\documentclass[10pt,twocolumn]{article}

\usepackage{graphicx}
\usepackage{amsfonts}
\usepackage{amsmath}
\usepackage{amssymb}
\usepackage{mathrsfs} % cursive letters
\usepackage{lipsum}
\usepackage{color}
\usepackage{siunitx}
\usepackage{authblk}
\usepackage[margin=2cm]{geometry}
\usepackage{mathtools, cuted}

\usepackage{bm}
\usepackage{amsmath}

\usepackage{color}
\definecolor{BDcolor}{rgb}{1,0,0}

\definecolor{Manuecolor}{rgb}{0,0,1}

\begin{document}

%\preprint{APS/123-QED}

\title{Coalescence in two-dimensional foams: a purely statistical process}% Force line breaks with \\
%\thanks{A footnote to the article title}%

\author{Emilie Forel $^1$, Benjamin Dollet $^2$, Dominique Langevin $^1$, Emmanuelle Rio$^1$\\
\small{$^1$ Univ. Paris Sud, Laboratoire de Physique des Solides, CNRS UMR 8502, Orsay, France\\
$^2$ LiPhy, Univ. Grenoble Alpes, CNRS, Saint-Martin d'H\`eres}\\
}

% \collaboration{MUSO Collaboration}%\noaffiliation

% \affiliation{
%  Third institution, the second for Charlie Author
% }%
% \author{Delta Author}
% \affiliation{%
%  Authors' institution and/or address\\
%  This line break forced with \textbackslash\textbackslash
% }%

%\collaboration{CLEO Collaboration}%\noaffiliation

\date{\today}% It is always \today, today,
             %  but any date may be explicitly specified
\twocolumn[
    \begin{@twocolumnfalse}
        \maketitle
           \begin{abstract}
While coalescence is ultimately the most drastic destabilization process in foams, its underlying processes are still unclear. To better understand them, we track individual coalescence events in two-dimensional foams at controlled capillary pressure. We obtain statistical information revealing the influence of the different parameters which have been previously proposed to explain coalescence. Our main conclusion is that coalescence probability is simply proportional to the area of the thin film separating two bubbles, suggesting that coalescence is mostly stochastic.
           \end{abstract}

    \end{@twocolumnfalse}]

%\pacs{Valid PACS appear here}% PACS, the Physics and Astronomy
                             % Classification Scheme.
%\keywords{Suggested keywords}%Use showkeys class option if keyword
                              %display desired

Liquid foams are essential to provide a pleasant texture to food and cosmetic products, to ensure thermal insulation or to serve as precursors for structural materials as concrete foams \cite{Stevenson2012}.
They are usually ephemeral, which can be very useful in many situations such as metallurgy, glass industry or water treatment, where the destruction of large volumes of foam after use is needed to extract the liquid for further processing. 
Among the different mechanisms leading to foam disappearance, coalescence, which is due to the rupture of the thin liquid film separating two neighboring bubbles, is the most efficient one.

However, coalescence remains uncontrolled, impeding a full exploitation of the potential of foams \cite{Rio2014,Colin2012,Langevin2012}. %The identification of the main mechanisms at the origin of film rupture is still under debate.
Several studies enlighten the importance of the liquid content (see references in \cite{Rio2014}), which decreases over time because of gravity, capillary drainage and evaporation. Foam then eventually reaches a dry enough state and collapses.
Nevertheless, the exact control parameters of coalescence, together with the appropriate mechanisms, remain elusive and are still debated.

A first scenario  states that it is induced by the capillary suction from the channels, called Plateau borders, formed at the intersection of three foam films.
their radius of curvature $R_{PB}$ in a foam stabilized by a solution of surface tension $\gamma$ sets the capillary pressure:
%\begin{equation} \label{Eq:capillary_pressure}
$P_c={\gamma}/{R_{PB}}$.
%\end{equation}
At equilibrium the film thickness is fixed by a balance between the disjoining pressure \cite{Israelachvili2011} and the capillary pressure. 
When the latter exceeds the equilibrium disjoining pressure, the thin films are prone to burst \cite{Khristov2002,Tcholakova2007}. Hence this scenario predicts a drastic upsurge of coalescence events below a critical $R_{PB}$.

In a second family of studies, the important parameter is the liquid fraction $\Phi$, which is the ratio between the volume of liquid within the foam and the volume of foam.
As discussed later, below a critical liquid fraction $\Phi_c$, avalanches of coalescence occur \cite{Carrier2003} because the shortage of liquid prevents any topological rearrangements necessary along foam aging \cite{Biance2011}.
 
A third scenario stems from the observation that foam film bursting seems stochastic \cite{Tobin2011}.
In such a scenario, the rupture probability of a film separating two bubbles increases with the film area, which scales as the square of the bubble radius $R_B^2$ \cite{Georgieva2009}.

To discriminate between these three mechanisms, it would be necessary to vary independently  $R_B$, $R_{PB}$ and $\Phi$. This is intrinsically difficult, since the radius of the Plateau borders is set by the bubble size and the liquid fraction as
\begin{equation}
R_{PB}^2 \propto \Phi R_B^2,
\label{eq:RPB}
\end{equation}
\noindent where $\Phi$ is a combination of $R_B$ and $P_c$ \cite{Cantat2013}.
Therefore, a foam can be considered as humid either because the capillary pressure is small, or because it contains small bubbles. 
Moreover, these parameters are not uniform in a given foam, since it can be polydisperse and exhibit gradients of liquid fraction leading to inhomogeneous values of $\Phi$, $R_B$ and $P_c$.
A last difficulty is that coalescence events are actually both rare and unpredictable \cite{Tobin2011,Rio2014}, so that  tackling the question requires an excellent statistics on such events. 
No former study varying one of these parameters independently was yet performed. Moreover, our study permits an unprecedented reliable statistical analysis and propose an original focus on the difference between average and local parameters.

In this Letter, we present experiments on 2D foams to record each coalescence event independently. 
A careful control of the capillary pressure together with the natural bubble growth due to coarsening, allows to vary the bubble size and the capillary pressure independently, leading to variable liquid fractions. 
This allowed us to identify a regime in which the best candidate among the three aforementioned scenarii is the stochastic one.

%======================================================================================
\begin{figure}[ht]
	\centering
    \includegraphics[width=8cm]{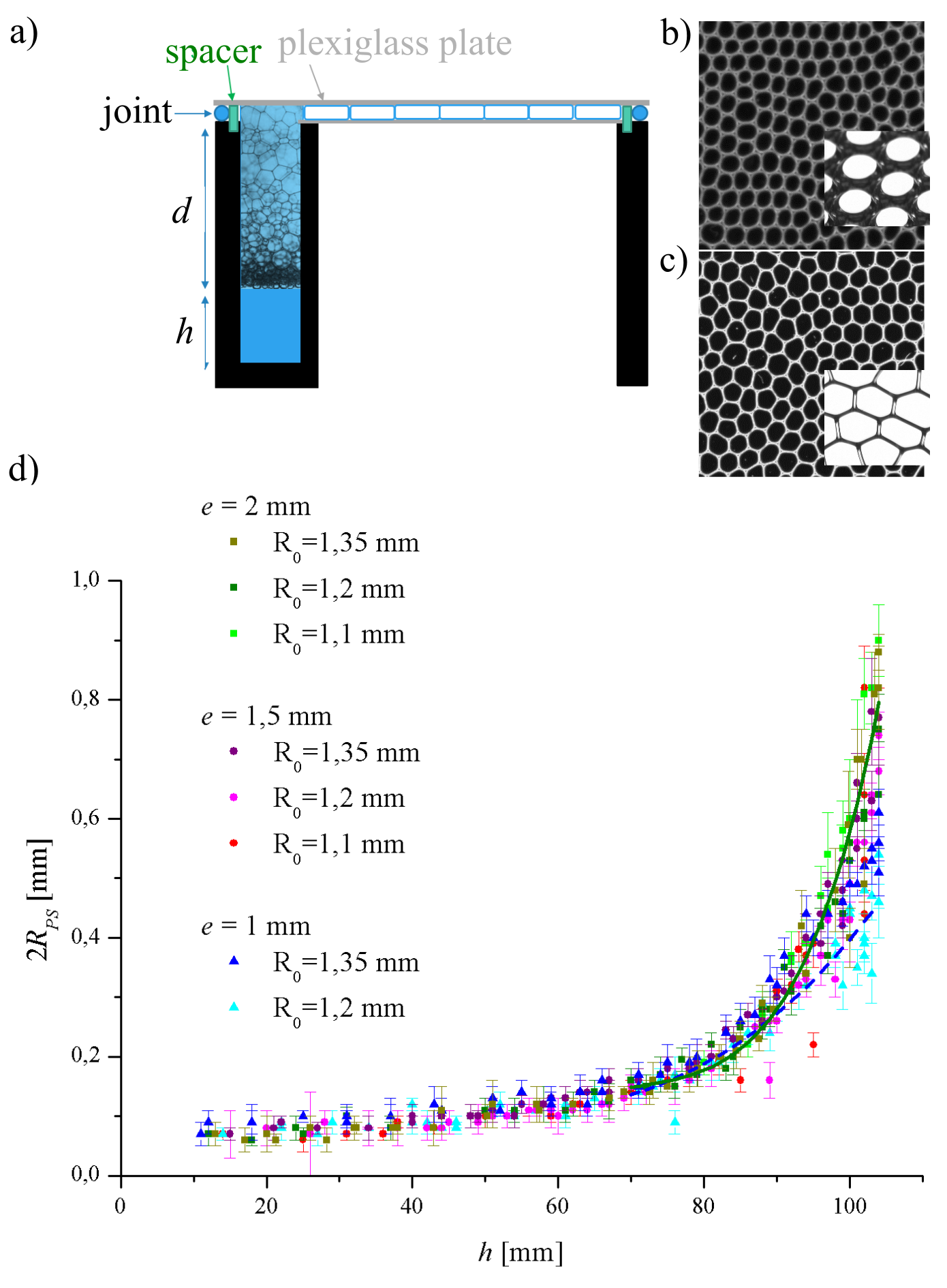}
	\caption{(a) Experimental setup: The liquid height $h$ allows to fix the capillary pressure in the 2D foam confined by the Plexiglas plates in a Hele-Shaw cell. (b) and (c) Top view of the bubbles for two different capillary pressures with a zoom at the scale of few bubbles in the inserts. (d) Evolution of the radius $R_{PS}$ of the pseudo Plateau borders with the liquid height. The lines are the fits used to extract $R_{PS}$ for each liquid height. For a spacer of 1 mm (dashed blue line), the equation is $ 3.3335 \times 10^{-3} h - 8.0883 \times 10^{-5} h^2 + 6.739 \times 10^{-7} h^3$ whereas for a spacer of 2 mm (solid green line), we used $7.09 10^{-3} h - 4.7568 10^{-4} h^2 + 1.3484 \times 10^{-5} h^3 - 1.6367 \times 10^{-7} h^4 + 7.2203 \times 10^{-10} h^5$
	}
    \label{fig:Setup}
\end{figure}
%======================================================================================

\begin{figure}[ht]
	\centering
    \includegraphics[width=8cm]{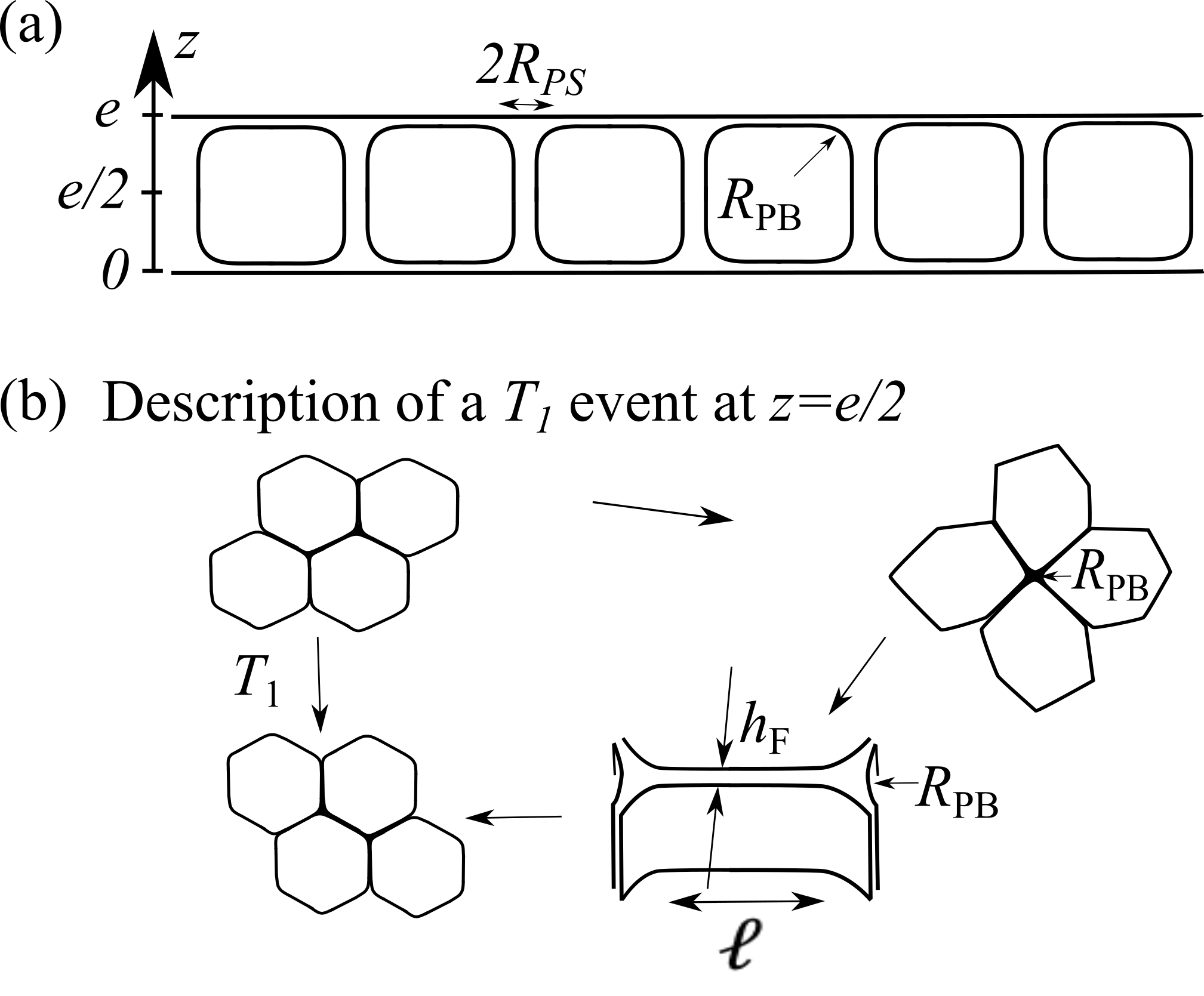}
	\caption{(a) Section of a 2D foam confined in a Hele-Shaw cell of thickness $e$. (b) %Horizontal section of the 2D foam at $z = e/2$ and $z = e$ showing respectively the structure in between the plates (where three bubbles meet in a channel of size $R_{PB}$) and at the contact with the top plate (where two bubbles meet the top plate in a channel of size $R_{PS}$). (c) 
	Sketch of the different steps along a $T_1$ event with the extraction of a film of length $\ell$ from a Plateau border of radius of curvature $R_{PB}$.
	}
    \label{fig:2DFoamStructure}
\end{figure}

The experimental setup, set on a heavy table to avoid vibrations, is inspired by the one proposed by Roth \textit{et al.} \cite{Roth2013}, as shown in Fig. \ref{fig:Setup}.
A 3D foam is created by bubbling air through a needle at the bottom of a reservoir filled with a solution of tetradecyltrimethylammonium bromide at a concentration of 5 g/L.
TTAB is a very common surfactant allowing to obtain long-lived foams.  In addition, it presents the advantage to be chemically stable along time.
The top of the reservoir is in contact with a Hele-Shaw cell of horizontal dimensions $10 \times 14$~cm$^2$, whose thickness is set by a spacer. 
A seal of width 1~cm prevents any leak or evaporation of the liquid during the experiment.
As the foam rises, a 2D foam invades the Hele-Shaw cell.
At equilibrium, a fixed liquid fraction profile is established in the 3D foam \cite{Maestro2013}.
Thus, a fixed height $h$ of liquid in the reservoir imposes a given value of $R_{PB}$ in the 2D foam (Fig.~\ref{fig:Setup}d), which in turns sets the capillary pressure in the 2D foam \cite{Roth2013}.
We characterize the foam liquid content by measuring the width $R_{PS}$ of the pseudo-Plateau borders, which are the channels at the junction between the top plate and two bubbles (Fig. \ref{fig:2DFoamStructure}).
A zoom on a few bubbles \cite{Forel2016} allows to measure accurately this parameter as a function of $h$  (Fig.~\ref{fig:Setup}b-c).
Fig.~\ref{fig:Setup}d shows that $R_{PS}$ increases continuously with $h$ and depends only a little on the bubble radius nor on the thickness $e$ separating the plates.
For the 36 experiments presented in this study, we fixed $e= 1$~mm or $e= 2$~mm.
The millimetric bubbles are close-packed so that more than 500 bubbles are present at the beginning of the experiment. 
A high-resolution camera (u-eye 1490, resolution $3840\times 2748$~pix$^2$) records the time evolution of the foam at a frequency of 1 frame per second.

%
%======================================================================================
%
%======================================================================================
%

\begin{figure}[ht]
	\centering
    \includegraphics[width=8cm]{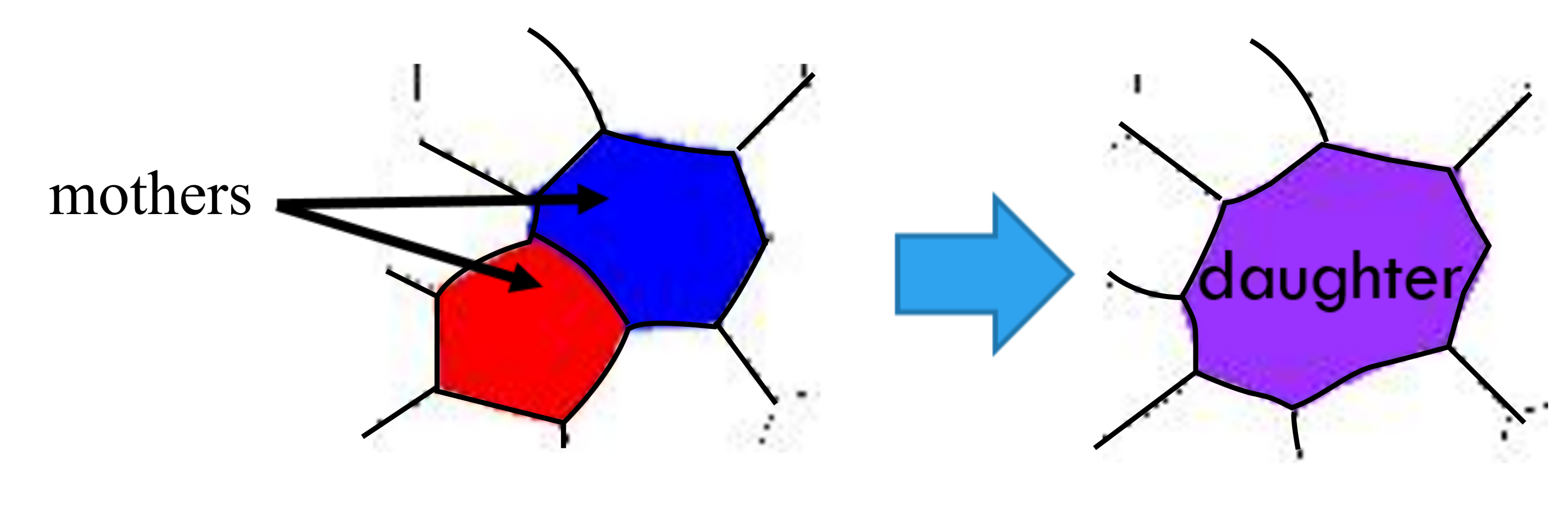}
	\caption{Detection of two mother bubbles and their common daughter. %(b) Count of the number of coalescence events along time during a single experiment performed for a liquid height $h=70$~mm and a cell thickness $e=1$~mm leading to a pseudo Plateau border $R_{PS}=68$~$\mu$m. In this experiment, we observed 188 coalescence events. 
	}
    \label{fig:figure2_nb_coal_vs_time}
\end{figure}

The foam evolution is automatically treated, which is necessary because of the rarity and unpredictability of the coalescence events. 
We adapted an existing home-made software \cite{Geraud2017}, which tracks each bubble between two successive images to detect individual coalescence events (SI A). 
For each of our 36 experiments, each individual event can be identified (Fig.~\ref{fig:figure2_nb_coal_vs_time}) and our software also allows to record the area $A_B$ of each bubble before and after coalescence as well as the length $\ell$ of each bubble side.
%======================================================================================
%======================================================================================
% Note that, at each time-step, each bubble can be considered as a bubble which does coalesce or not.
Thus, we recorded $2.9 \times 10^7$ bubbles (accounting for their multiple occurrence in different images of a given movie), among which 5668 did coalesce.
This enlightens the rarity of the events, and the importance of automatic treatment to get sufficient statistics.

To discriminate between the possible coalescence mechanisms, we measured the parameters relevant to the different scenarii.
%For each experiment, we measure the value of $h$ by direct visualization through a lateral window.
%We fitted the curve  observed in Fig.~\ref{fig:Setup}b for each value of the spacer by a polynomial function of degree 5 to obtain a direct relationship between $h$ and $R_{PS}$.
%The film area between separating the two mother bubbles is obtained for each bubble by multiplying the thickness separating the glass plates $e$ by the length $\ell$ of the bubble side.
%To compare our results to the scenario proposed by Biance \textit{et al} \cite{Biance2011}, we will show in the following that the right parameter is the ratio ${R_{PS}}/{\ell}$.
%======================================================================================
\begin{figure}[ht]
	\centering
    \includegraphics[width=8cm]{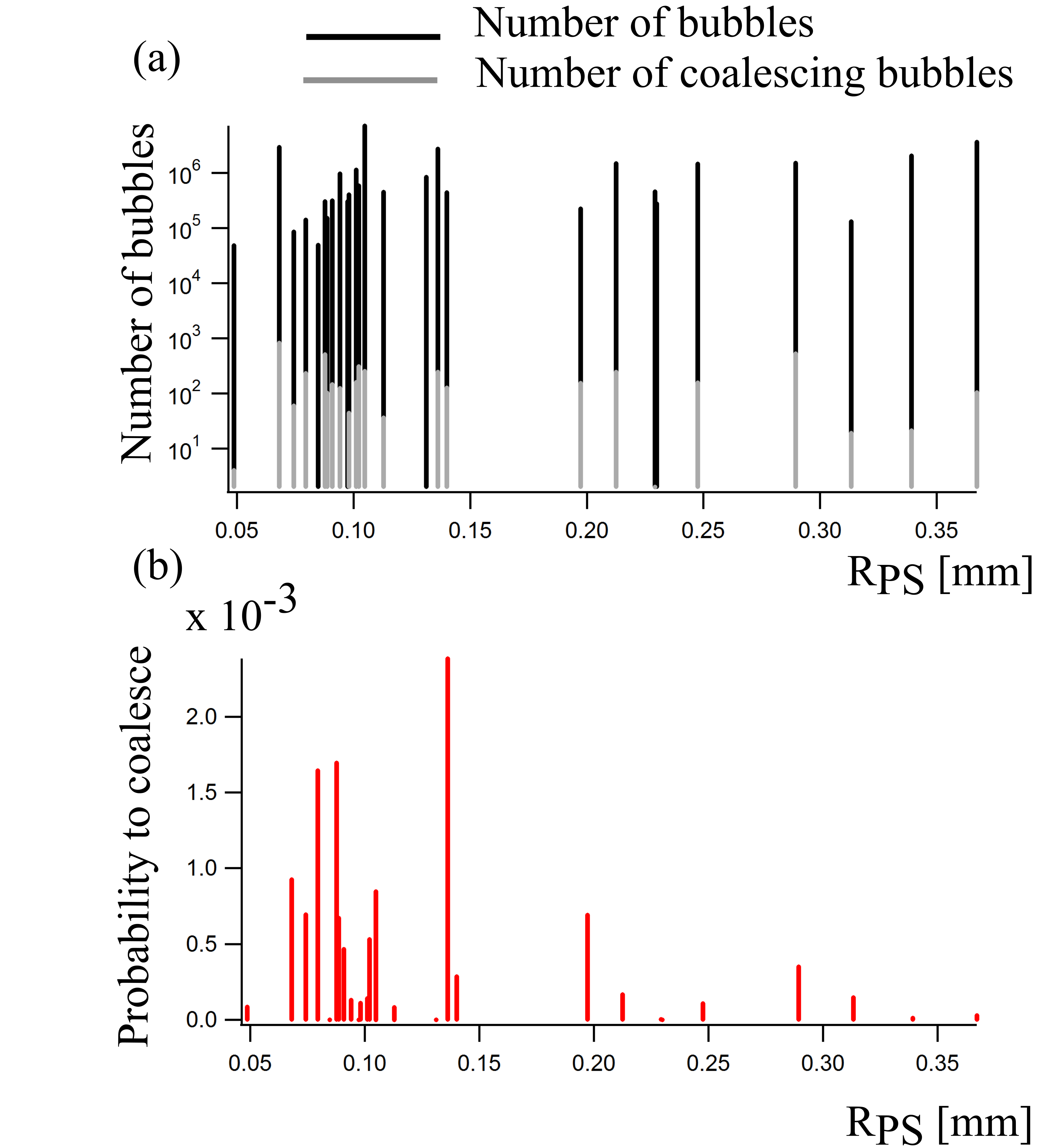}
	\caption{For each experiment, at fixed $R_{PS}$, plot of (a) the number of observed bubbles and the number of coalescence events, and (b) the probability of a bubble to coalesce.
	}
    \label{fig:figure_3_nb_proba_vs_rPB}
\end{figure}
%======================================================================================
%In this letter, we explore the impact of each of the three aforementioned parameters on foam coalescence.
In Figs.~\ref{fig:figure_3_nb_proba_vs_rPB}a, \ref{fig:figure_5_nb_proba_vs_phis}a and \ref{fig:figure_4_nb_proba_vs_Rb}a, we plot, for each value of the tested parameter, the total number of bubbles that we observed, $N_o$ (in black) and the number of these bubbles which coalesce, $N_c$ (in grey).
% Note that the y-axis is in logarithmic scale, which is another signature of the rarity of the coalescence events.
The ratio $N_c/N_o$ is the coalescence probability between two images. It is plotted in Figs.~\ref{fig:figure_3_nb_proba_vs_rPB}b, \ref{fig:figure_5_nb_proba_vs_phis}b and \ref{fig:figure_4_nb_proba_vs_Rb}b.

To test the scenario of the critical capillary pressure, we studied the influence of the pseudo-Plateau border radius $R_{PS}$.
For each experiment, we measure the value of $h$ by direct visualization through a lateral window.
We fitted the curve  observed in Fig.~\ref{fig:Setup}d for each value of the thickness to obtain a direct relationship between $h$ and $R_{PS}$ (see caption of Fig.~\ref{fig:Setup}).
In Fig.~\ref{fig:figure_3_nb_proba_vs_rPB}, each value of $R_{PS}$ corresponds to a given experiment.
We observe that, whatever the size of the Plateau borders, coalescence events occur. 
Moreover, the probability to coalesce exhibits no clear tendency.
This shows that, in our configuration, the capillary pressure is not the right parameter to describe the coalescence.

We now study the scenario of a critical liquid fraction proposed by Biance \textit{et al} \cite{Biance2011}.
According to these authors, below a critical liquid fraction, the films break during topological rearrangements called T$_1$ events.
Indeed, for bubbles to swap neighbors, a new soap film is created. 
Owing to dynamic effects, this film is thicker during its extension following its creation than at equilibrium. In a too dry foam, the quantity of liquid available in the Plateau borders is too small to feed this extending film, and it pops.
In 2D foams, the same mechanism can be invoked.
During a $T_1$ event, the film is generated from a channel formed by the transient intersection of four bubbles (Fig.~\ref{fig:2DFoamStructure}c). 
%This transient Plateau border has a radius $R_{PB}=R_{PS}/\sqrt{3}$ \cite{Schimming2017} so that its section can be evaluated as $\left(4-\pi \right) R_{PB}^2$. It thus contains a volume of liquid of $\left(4-\pi \right){R_{PS}^2} e/3$, where $e$ is the vertical height of the film \BD{a-t-on besoin d'introduire cette nouvelle notation, n'a-t-on pas $e = e_F$~?)} \Manue{Pas vraiment. Les deux ne sont pas egaux mais comme ca s'annule, je suis d'accord pour qu'on passe cette subtilite sous le tapis.}.
This transient Plateau border has a radius $R_{PB}=R_{PS}/\sqrt{3}$ \cite{Schimming2017} so that its section scales as $R_{PB}^2$. 
It thus contains a volume of liquid scaling as ${R_{PS}^2} e/3$, where $e$ is the vertical height of the film.
Moreover, a new film contains a volume $e h_F \ell$ of liquid, where $h_F$ is the transient thickness of the extending film.
As suggested by Biance \textit{et al.} \cite{Biance2011}, we propose that the film thickness is fixed by the velocity of the rearrangements $V$, i.e. the typical extension velocity of a film after its creation in a $T_1$ event, through Frankel's law,
\begin{equation} \label{Eq:Frankel}
h_F=1.84 R_{PB} \rm{Ca}^{2/3} ,
\end{equation}
with the capillary number ${\rm{Ca}} = \eta V/\gamma$ with $\eta$ the liquid viscosity.
The film is expected to burst during the rearrangements as soon as the Plateau border volume is comparable to the film volume, i.e. $e h_F \ell \approx  e R_{PS}^2$
%During a small variation of the film length $\rm{d}\ell$, the variation of the volume of the film is balanced by the variation of the volume of the Plateau border
%
%\begin{equation}
%h_F \mathrm{d}\ell e = \frac{2}{3} R_{PS}e \mathrm{d}R_{PS}
%\end{equation}
%
which, combined with (\ref{Eq:Frankel}), yields to a critical value of ${R_{PS}}/{\ell}$ scaling as
%\begin{equation}
${R_{PS}}/{\ell} \approx \rm{Ca}^{2/3}$.
%\end{equation}
%With a surface tension $\gamma=36$ mN/m, a viscosity $\eta=10^{-3}$ Pa.s and a typical velocity of rearrangements $V=0.1$ m/s \cite{Biance2011}, we compute ${R_{PS}}/{\ell}=0.037$.
The right parameter to account for this mechanism is finally ${R_{PS}}/{\ell}$.

We thus plot the probability of a bubble side to break versus the value of ${R_{PS}}/{\ell}$.
Two families of bubbles can exist.
For very small bubbles in a rather wet foam, the length $\ell$ of their frontier with a neighboring bubble is smaller than $2R_{PS}$ and the bubbles are spherical.
For larger bubbles and/or drier foams, a flat film appear because ${R_{PS}}/{\ell} < 1/2$.
Given the excellent sampling of bubble size, the second family of bubbles is well described with a continuous variation of ${R_{PS}}/{\ell}$ between 0 and 0.5.
The sampling is not as good for the first family, but this is unimportant since these wet bubbles were never observed to coalesce.
Finally, the number of observed coalescence events observed versus ${R_{PS}}/{\ell}$ gives the very smooth curve in Fig.~\ref{fig:figure_5_nb_proba_vs_phis}a, which lends confidence to the probability shown in Fig.~\ref{fig:figure_5_nb_proba_vs_phis}b.
In particular, we mention that, for each value of this parameter, more than $10^3$~bubbles have been observed.
%======================================================================================
%
\begin{figure}[ht]
	\centering
    \includegraphics[width=8cm]{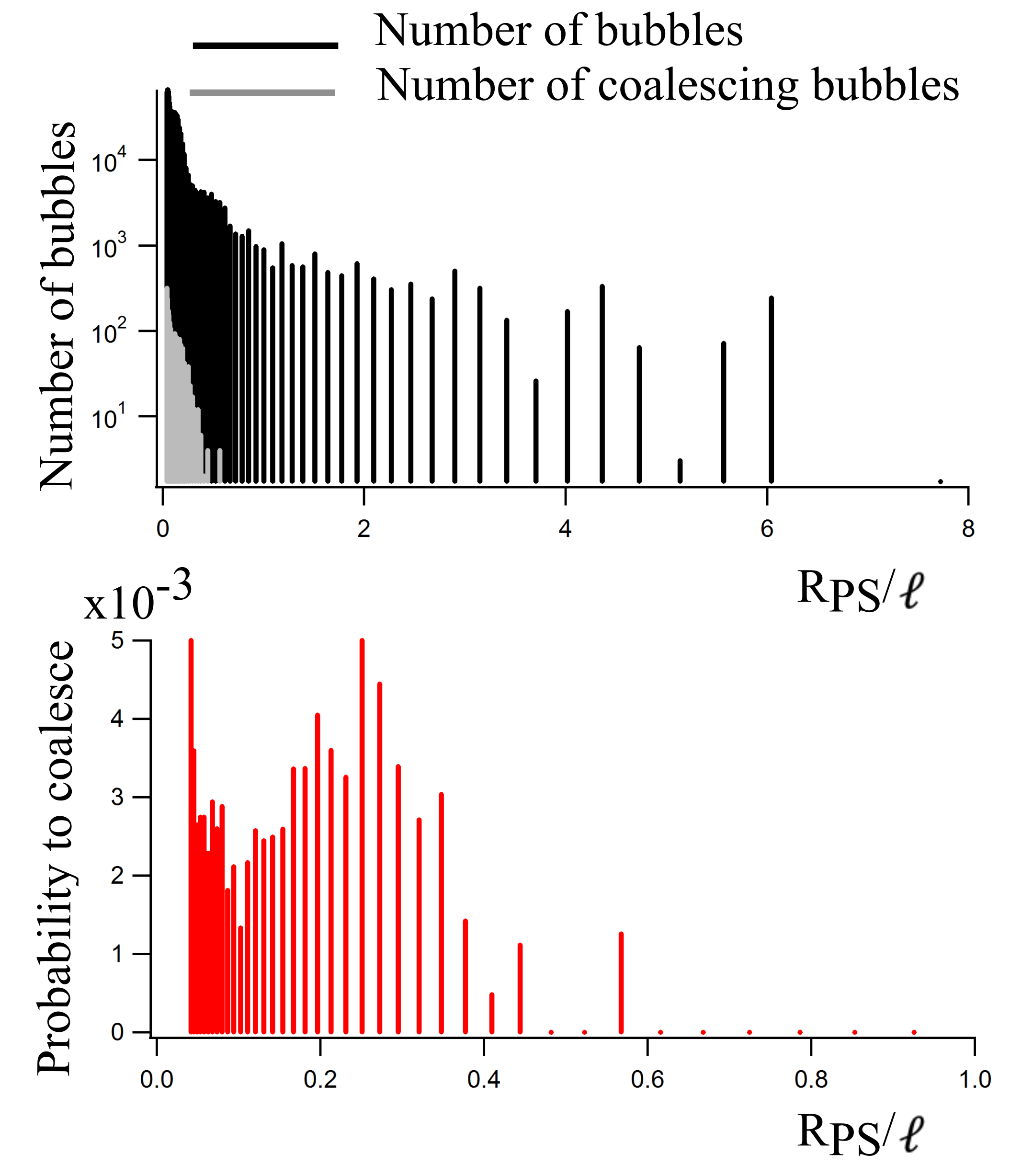}
	\caption{Plot of (a) the number of observed bubbles and of coalescence events and (b) the probability of a bubble to coalesce, as functions of ${R_{PS}}/{\ell}$. Note that the $x$-axis is different in both plots.
	}
    \label{fig:figure_5_nb_proba_vs_phis}
\end{figure}
%======================================================================================
The first observation is that no coalescence events are observed for ${R_{PS}}/{\ell}$ larger than 0.5 whereas
we observe a nonzero probability for the bubble to coalesce for all values of ${R_{PS}}/{\ell}$ smaller than 0.5 (Fig.~\ref{fig:figure_5_nb_proba_vs_phis}a).
Nevertheless, we observe no systematic variation of the probability for a bubble side to rupture in this parameter range (Fig.~\ref{fig:figure_5_nb_proba_vs_phis}b).
This naturally excludes the $T_1$ scenario from the mechanisms triggering to coalescence.
%======================================================================================
\begin{figure}[ht]
	\centering
    \includegraphics[width=8cm]{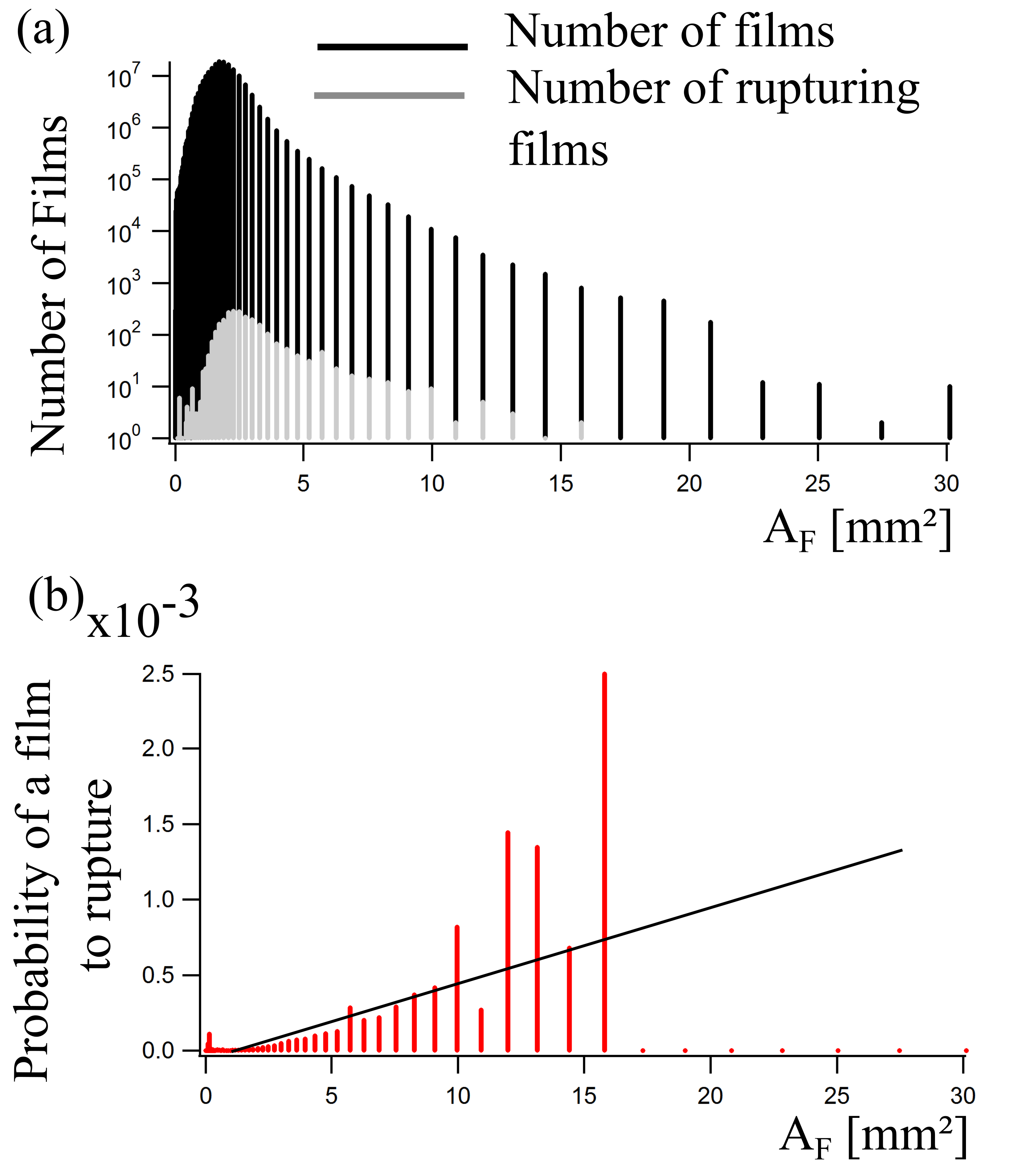}
	\caption{Plot of (a) the number of observed films and of coalescence events and (b) the probability of a film to break, as functions of the film area. The black solid line is a guide for the eye.
	}
    \label{fig:figure_4_nb_proba_vs_Rb}
\end{figure}
%======================================================================================

The last scenario is the stochastic one, in which the relevant parameter is the area $A_F$ of the film separating the two mother bubbles.
This area is given by $A_F = e\ell$ where $e$ is the thickness separating the glass plates and $\ell$ the length of the film.
Fig.~\ref{fig:figure_4_nb_proba_vs_Rb} shows the influence of $A_F$ on the coalescence events.
The film area grows continuously over time, because the mean bubble area increases due to both coarsening and coalescence.
This explains the excellent sampling observed in Fig.~\ref{fig:figure_4_nb_proba_vs_Rb}, at least for the bubbles of area under 20 mm$^2$.%radius under $2$ cm.
If coalescence were purely stochastic along the film area, we would observe a linear increase of the rupture probability with the film area.
As shown in Fig.~\ref{fig:figure_4_nb_proba_vs_Rb}b, such a linear increase is actually observed, except for very small and very big films.  
The scatter observed  for bigger films corresponds to big bubbles for which the statistics worsens.
The scatter may be due to lack of statistics: there was less than ten coalescence events (for small films) or bubbles (for large films) per bin.
No coalescence was detected for the largest films ($A_F > 16$~mm$^2$), but only few of them could be detected reliably. With this word of caution,
% Bigger films correspond to big bubbles and we would like to note that the statistics is poor for large bubbles since only one or two big bubbles are observed for each bubble size \BD{(je ne suis pas s\^ur que les deux barres d'histogramme soient statistiquement significatives. Ne pr\'er\`eres-tu pas les enlever, et \'ecrire qu'on ne garde que les intervalles avec plus de $10^2$ bulles~?)} \Manue{Je suis un peu perdue. Est ce qu'il y a encore des barres qui ne vont pas ?}.
% Indeed, at the end of the experiment, the bubble size can be comparable to half the size of the cell.
% Moreover, the bigger bubbles are often in contact with the borders so that we exclude them from our data.
% Together with the lack of statistics, this may explain why no coalescence events are observed for the larger bubbles.
these results are then compatible with a stochastic rupture of the foam films, except for very small bubbles.

In conclusion, we developed an experiment allowing to detect single coalescence events in a 2D foam by image analysis.  
We were able to control the capillary pressure and to measure the foam film size, as well as the ratio ${R_{PS}}/{\ell}$ between the size of the pseudo-Plateau borders and the length of the ruptured film. 
This allowed us to fully characterize almost 6000 coalescence events and to obtain unprecedented statistics concerning the impact of the capillary pressure, the foam film size and the ratio ${R_{PS}}/{\ell}$ on the probability of a bubble to coalesce. 
Our conclusions are that neither the capillary pressure nor the liquid fraction trigger coalescence in our explored range of parameters. 
The probability for a film to rupture is rather linear with the film area, in accordance with a purely statistical scenario , except for very small films.
The reason why a critical capillary is sometimes observed in the literature could be the explored range of $R_{PS}$.
The capillary pressures indeed varies in between 100 and 700 Pa in our experiment, which is of the order, but smaller than the critical capillary pressure observed by Khristov \textit{et al} \cite{Khristov2002}. 
It is likely that above a critical capillary pressure, the films may rapidly rupture, while below this pressure, film rupture is stochastic and less frequent.

To generalize the impact of the foam films area in 3D foams, a study similar to the one described here, i.e. at the scale of each bubble, would be necessary to overcome the inhomogeneities in size and liquid fraction inherent to aging bubbles.
Such experiments are of course very challenging; nevertheless, our results suggest that the statistical rupture of a film should remain important for coalescence in 3D foams as well. 

\section*{Acknowledgements}
We acknowledge funding from ESA (MAP Soft Matter Dynamics and contract 4000115113) and CNES (through the GDR MFA).
We are grateful to Isabelle Cantat and Anne-Laure Biance for fruitful discussions, and to Gabriel Le Doudic and Louis Rioux for performing some of the experiments. 
The experiments necessitate the control of the capillary pressure provided by the contact between the 2D and the 3D foam and we are thankful to Douglas Durian for providing the plans of his own cell.

\bibliographystyle{unsrt}
\bibliography{Biblio}

\end{document}